\input phyzzx
\centerline{On the $\Theta$-term in electrodynamics}
\vskip .5cm
\centerline{Pawel O. Mazur$^1$\footnote{*}{E-mail: mazur@psc.psc.sc.edu},
Andrzej Staruszkiewicz$^2$\footnote{**}{E-mail: astar@thrisc.if.uj.edu.pl}}
\vskip .5cm
\centerline{$^1$ Department of Physics and Astronomy,}
\vskip .1cm
\centerline{University of South Carolina, Columbia, SC 29208, USA}
\vskip .5cm
\centerline{$^2$ Institute of Physics, Jagellonian University}
\vskip .1cm
\centerline{Reymonta 4, 30-059 Krakow, Poland}
\vskip .5cm
\centerline{Abstract}
\vskip .5cm
 The term $\Theta{\epsilon^{\mu\nu\rho\sigma}}F_{\mu\nu}F_{\rho\sigma}$,
when added to the electromagnetic Lagrangian\par
$-{1\over 16\pi}F^{\mu\nu}F_{\mu\nu}$, does not change the signature
of the Lagrangian. Actually, \par 
it increases the part with 
negative kinetic energy term
at the spatial infinity. \par
For this reason it does not change the conclusion,
that at the spatial infinity \par
the magnetic part of the electromagnetic field should be absent.
\vskip 1cm
\centerline{PACS numbers: 12.20.Os ,  11.10.Ij}
\vskip .5cm
\centerline{Keywords: Dirac magnetic monopoles, infrared behavior, QED}
\vfill
\eject
\par
The best established conservation law in physics is the electric charge
conservation. The same is true for the universality of electric charge,
i. e., the equality of the absolute magnitude of electric charge
of an electron and a proton. The most remarkable fact in physics
is the quantum nature of an electric charge. This fact was well
established even before the discovery of quanta of energy.
In 1909 Einstein brought to the broader audience another remarkable
fact, observed earlier by Jeans, that the magnitude of the electric
charge squared $e^2$ has the same physical dimension as $hc$, where $h$
is the Planck constant introduced just few years before. The magnitude
of the ratio $e^2\over hc$ was then estimated to be of an order of
$10^{-3}$. Einstein has proposed that the same theoretical framework which
will have a constant $e^2$ included in its mathematical structure will
have as a consequence the quantum theory of radiation, and, therefore, 
the Planck constant $h$ will have been `explained'. 
In other words, once $e^2$ and the fine
structure constant $\alpha={e^2\over \hbar c}$ are given then $h$ would
have been established as a secondary constant of Nature. This was not
what have happened historically, as we well know now. Meanwhile the quantum
theory of radiation was established but it was also recognized as an
incomplete theoretical scheme. This is precisely due to the
remarkable experimental fact of the electric charge quantization $Q=Ne$. 

Since Gauss we know that the electric charge `resides 
at spatial infinity', as we would describe it in our modern language. 
On the other hand special theory of relativity tells us 
that spatial infinity where electric charge 
resides is a dynamic concept as an electric charge exists
for the eternity of time allowed to it at spatial infinity.  
Physically, a signal propagating from or to spatial infinity 
takes an infinite duration of time. 
This can be formally understood from the observation
that the Gauss law is valid in every Lorentz frame. 
The phenomenological theory of the electric charge 
proposed by one of us [1,2] contains the only constant 
of Nature which is relevant to the problem of
the quantum nature of electric charge, which is $e$, the
magnitude of the electronic charge. Staruszkiewicz proposed 
a decade ago [1] that the closed dynamical system 
which contains electric charge must
necessarily contain infrared photons which carry information about
electric charges emitting them so they could be observed
at spatial infinity. 

It is well known that spatial infinity 
of the Minkowski spacetime is the timelike $2+1$-dimensional 
de Sitter hyperboloid. This is what is needed for the purposes 
of doing quantum field theory because such a manifold 
has a well defined Cauchy surface. 
Quantum mechanics of the electric charge is
the quantum field theory of the phase field 
$S(x)$ defined on the de Sitter spatial infinity [1]. 
In [2] and below the phase field is denoted $e(x)$ for obvious 
reasons. 

Electromagnetic field at the spatial infinity is described completely
by two homogeneous of degree zero solutions of the d'Alembert equation
[1,2]. They are defined as follows. At the spatial infinity the potential
$A_{\mu}(x)$ must be homogeneous of degree $-1$:

$$A_{\mu}(\lambda x) = {\lambda}^{-1}A_{\mu}(x)   ,  \eqno(1)$$
\noindent
for all $\lambda>0$ [1,3].
\par
Using the Maxwell equations and the above homogeneity condition one finds that

$$x^{\mu}F_{\mu\nu}(x) = \partial_{\nu}e(x)  ,  $$

$${1\over 2}{\epsilon^{\mu\nu\rho\sigma}}x_{\nu}F_{\rho\sigma}(x) =
\partial^{\mu}m(x)  .   \eqno(2)$$
\noindent
\par
These equations can be solved with respect to $F_{\mu\nu}$,
which shows that the functions $e(x)$ and $m(x)$ determine $F_{\mu\nu}(x)$
completely; $e(x)$ is the electric part of the field while $m(x)$
is the magnetic part. It was shown in [1,2] that

$$-{d^4}xF_{\mu\nu}F^{\mu\nu} = 2{d{\zeta^0}\over {\zeta^0}}\sqrt{g}
{d^3}{\zeta}
(g^{ik}{\partial_i}e{\partial_k}e - g^{ik}{\partial_i}m{\partial_k}m)  .
\eqno(3)$$
\noindent
The metric on spatial infinity $g_{ik}$ is defined in an obvious way

$$g_{ik} = ({\zeta^0})^{-2}g_{\mu\nu}{{\partial}x^{\mu}\over \partial{\zeta^i}}
{{\partial}x^{\nu}\over \partial{\zeta^k}},  i,k=1,2,3 .   \eqno(4)$$
\noindent
The coordinates covering spatial infinity,
${\zeta}^0=\sqrt{-xx}{\rightarrow +\infty}$, are the hyperspherical
coordinates [2]:

$$\eqalign{
x^0& = {\zeta^0}sinh{\zeta^1},\cr
x^1& = {\zeta^0}cosh{\zeta^1}sin{\zeta^2}cos{\zeta^3},\cr
x^2& = {\zeta^0}cosh{\zeta^1}sin{\zeta^2}sin{\zeta^3} ,\cr
x^3& = {\zeta^0}cosh{\zeta^1}cos{\zeta^2} .\cr
}
\eqno(5)$$
\noindent
\par
The Lagrangian density (3) is seen to be a difference of two identical
Lagrangian densities.
The part with the right sign, 
giving rise upon quantization to a positive definite inner product, 
is called electric. 
The part with the wrong sign is called magnetic. 
It is seen that the magnetic part enters the total Lagrangian with
the negative sign. This is unphysical and probably explains nonexistence
of magnetic monopoles [6,2]. 
We hold it selfevident that the sign of the Lagrangian
is physically important and that the wrong sign implies the existence
of negative norm states. One may keep them, 
but then one is not working in the framework of quantum mechanics [5].
\par
We wish to note, that the addition of the term
$\Theta{\epsilon^{\mu\nu\rho\sigma}}F_{\mu\nu}F_{\rho\sigma}$ to
the Lagrangian [4] does not change this conclusion, simply because it does not
change the signature of the Lagrangian treated as a quadratic form:

$${d^4}x{\bigl(-{1\over 16\pi}F_{\mu\nu}F^{\mu\nu} +
\Theta{\epsilon^{\mu\nu\rho\sigma}}F_{\mu\nu}F_{\rho\sigma}\bigr)} =
{1\over 8\pi}{d{\zeta^0}\over {\zeta^0}}{\sqrt{g}}{d^3}{\zeta}
{\bigl(g^{ik}{\partial_i}e'{\partial_k}e'
- g^{ik}{\partial_i}m'{\partial_k}m'\bigr)}  ,    \eqno(6)$$
\noindent
where

$$e' = ecosh{\gamma} - msinh{\gamma} , $$
$$m' = esinh{\gamma} + mcosh{\gamma} .   \eqno(7)$$

\noindent
The parameter $\gamma$ is defined by the relation:
$sinh{2\gamma} = {8\pi\Theta}$.

Therefore, the asymptotic Lagrangian at the spatial infinity, calculated as
above (Eq.(3)), but including the $\Theta$ term (Eq.(6)), will also
be a difference of two identical Lagrangians, one having necessarily
the `wrong' sign. This means, that the argument against the existence
of magnetic monopoles [6] given at [2] is not affected by the $\Theta$ term.
It was also shown in [1,2] that the electric charge $Q$ 
is always quantized in the units of electronic charge $e$. 
Magnetic monopoles (if they existed) would possibly carry a fractional 
electric charge [4].     
Hence, nonexistence of magnetic monopoles 
is compatible with the quantization of electric charge [1,2]. 
\vskip .1cm

\centerline{Aknowledgements}

 This research was partially supported by NSF grant to University of South
Carolina (P.O.M.), and by Polish KBN grant and University of South Carolina
(A.S.). We would also like to acknowledge the warm hospitality of the
Jagellonian University (P.O.M.) and University of South Carolina (A.S.) during
the time this work was first written up (September 1996). 
\vfill
\eject
\centerline{References}

\item{ 1.}   A. Staruszkiewicz, Ann. Phys. (N.Y.) {\bf 190} (1989) 354.
\item{ 2.}   A. Staruszkiewicz, in: {\it ``Quantum Coherence and Reality''},
Yakir Aharonov's Festschrift, eds. J.S. Anandan and J.L. Safko
(World Scientific, Singapore 1994).
\item{ 3.}   J.-L. Gervais, and D. Zwanziger, Phys. Lett. {\bf 94B} (1980) 389.
\item{ 4.}   E. Witten, Phys. Lett. {\bf 86B} (1979) 283.
\item{ 5.}   S. W. Hawking, and S. F. Ross, Phys. Rev. {\bf D52} (1995) 5865.
\item{ 6.}   P. A. M. Dirac, Proc. Roy. Soc. (London) {\bf A133} (1931) 60.

\end